\documentstyle[12pt]{article}
\catcode`\@=11
\def\slash{\@ifnextchar[{\@slash}{\@slash[\z@]}}
\def\@slash[#1]#2{\setbox\z@\hbox{$#2$}\@tempdima\wd\z@\box\z@%
\@tempdimb#1 \advance\@tempdimb-\@tempdima \kern\@tempdimb
\hbox to\@tempdima{\hss\@makeslash\hss}}
\def\@makeslash{$/$}			
\catcode`\@=12
\def\FMslash{\slash}

\newcommand{\be}{\begin{equation}}
\newcommand{\ee}{\end{equation}}
\newcommand{\ba}{\begin{eqnarray}}
\newcommand{\ea}{\end{eqnarray}}
\newcommand{\bas}{\begin{eqnarray}}
\newcommand{\eas}{\end{eqnarray}}
\def\AA{{\cal A}}
\def\DD{{\cal D}}
\def\HH{{\cal H}}
\def\II{{\cal V}}

\def\dd{{\rm d}}
\def\tr{{\rm tr}}
\def\half{{1 \over 2}}
\def\Zset{{\bf Z}}

\def\Cset{{\bf C}}
\def\unity{{\bf 1}} 
\def\ie{ie.}
\def\herm{Hermitian}
\def\aherm{anti-Hermitian}
\def\abel{Abelian}
\def\nabel{non-Abelian}
\def\nondyn{nondynamical}  
\def\eq{Eq.} 

\begin{document}
\begin{titlepage}

\begin{flushright}
 UU-ITP-26/96 \\
 hep-th/9612027
\end{flushright}
\vskip 0.5truecm

\begin{center}
{\bf \Large \bf Dimensionally Reduced Yang--Mills 
Theories   \\
\vskip 0.2cm
in Noncommutative Geometry \\ 
}
\end{center}

\vskip 0.7cm
\begin{center}
{\bf Jussi Kalkkinen$^{*}$} \\
\vskip 0.6cm
{\it Department of Theoretical Physics,
Uppsala University \\
P.O. Box 803, S-75108, Uppsala, Sweden  \\}
\end{center}

\vskip 1.5cm
\rm
\noindent
We study a class of noncommutative geometries that 
give rise to dimensionally reduced Yang--Mills 
theories. 
The emerging geometries describe sets of
copies of an even dimensional manifold. Similarities 
to the D-branes in string theory are discussed.
\vfill

\begin{flushleft}
\rule{6.2 in}{.007 in} \\
$^{*}$  \small E-mail: 
 jussi@teorfys.uu.se \\
\end{flushleft}

\end{titlepage}
\vfill\eject
\baselineskip 0.65cm

\section{Introduction}

Connes' theory of noncommutative geometry (NCG) has enabled 
physicists to study configuration spaces with a geometry 
decisively more general than that of the usually considered 
differentiable  manifolds \cite{connes}. 
In these constructions geometry gives naturally 
rise to a characteristic field theory defined on the 
considered noncommutative space. Not all field theories, 
however, can yet be obtained this way. 
Those that can, 
include the classical field theory formulations of the 
General Relativity, the Standard Model, 
grand unified theories and 
some space--time supersymmetric theories 
in four dimensions \cite{ncgs,froeh_gut,cham_susy}. 
These studies have lead to 
the understanding of the geometric origin of the Higgs' 
scalar field in the Standard Model. 

All of these NCG's are modifications of the idea of 
expanding a manifold into a collection of its copies
(called p-branes in the following) in 
a nontrivial way. In this context,
the NCG origin of a field theory strongly
constrains the form of the scalar potential
\cite{cham_pot}. In this article we shall consider a 
particular subclass of these NCG models, 
namely those that give 
rise to a dimensionally reduced Yang--Mills theory. 
The idea of dimensional reduction may seem foreign to 
NCG at the first sight but a closer look at it 
leads to a natural generalization of an embedding of 
geometric objects into a larger, partly compactified space. 
The compactified directions turn out to 
appear as relations between 
the embedded objects -- no reference 
to an actual embedding space is necessary. 

This work was originally 
motivated by the apparent similarity of the 
above mentioned NCG's and the
D(irichlet) p-branes that appear in string theory. 
These are in $D$ dimensions embedded $p+1$-dimensional 
surfaces on which the boundaries of the 
open string world sheets are constrained \cite{polchinski}. 
The vanishing of the 1-loop beta-functions of the open 
strings is equivalent to the fact that the classical 
equations of motion of the D-branes (coming from a 
Born--Infeld action) are satisfied. The emerging
low energy effective field theory is to the 2nd order 
in the field strength a dimensionally reduced 
Yang--Mills theory. The similarity of 
the topologies of the two systems leads one to ask whether 
there indeed is a NCG capable of producing the expected low 
energy field theory and the geometric relations between 
the D-branes. 

In this work we shall study NCG's that give rise 
to these field theories, but  describe a 
geometry -- in the sense of measuring distances -- 
that does not coincide with the D-brane results. 
These problems  were already pointed out in 
Ref.~\cite{douglas}. In  what follows we study 
this particular class of NCG's on its own right, and point 
out many of the similarities
to and differences from the D-branes. An other appearance of 
NCG in string theory was considered in Ref.~\cite{shenker}, 
where the M-theory membranes were studied from the 
NCG point of view.
After the completion of the 
present work we received Ref.~\cite{utah} where similar 
issues were considered.   

\section{Connes' NCG}

Let us start by considering a collection of copies of a 
Riemannian manifold  $(M,g)$ in NCG \cite{connes,varilly}. 
In Riemannian geometry one would embed the 
copies of $M$ in a bigger space in order to address 
questions 
concerning their mutual relations. In NCG this is not 
necessary, but the formalism is particularly  
well suited for studying geometry in spaces that are of the 
form $M \times {\Zset}_n$, ${\Zset}_n$ 
being a discrete set of points. 
The basic data needed for this   
are a K-cycle $(\HH,\DD,\Gamma)$ on a *-algebra $\AA$. 
In the present case $\AA$ can be taken to be the algebra of 
$M_n({\Cset})$-valued (smooth) functions on $M$, 
$\HH$ the Hilbert space of square integrable 
spinors on $M$ tensored with the $n$-dimensional  
representation space of 
$M_n({\Cset})$, $\Gamma$ the Clifford grading and 
\bas
{\cal D} = \FMslash{\partial} 
\otimes {\unity} + \Gamma \otimes N,
\eas
where $N \in M_n({\Cset})$ is an \aherm~matrix 
(with a vanishing diagonal) of dimension mass. 

\subsection{The metric}

The measurement of distances will turn out to be the 
distinguishing factor between our NCG models and 
the D-brane models, so we shall need to consider 
it at some length.  

In the case $n=1$ we are 
considering a single copy of $M$. In 
this case the geodesic distance between two points 
$p,q \in M$ can be calculated from 
\ba
d(p,q) = \sup_{a \in \AA} \Big\{  | a(p) - a(q) | ;~~ {} 
 \parallel  [{\cal D}, a  ]  \parallel \leq 1 \Big\},   
\label{metriikka0}
\ea
where the norm $\parallel ~ \parallel$ is the  
operator norm in ${\rm End} ~\HH$. 
This metric is the same as the one found by 
looking for shortest paths 
using the metric tensor $g$ \cite{connes}. 
The term $a(p) = \chi(a)$ should be seen as a 
character of the algebra $\AA$, \ie~a homomorphism 
{}from $\AA$ to complex numbers. In the case $n>1$ 
we consequently can speak about points of a 
noncommuting manifold with a function algebra $\AA$ if we 
know what the characters are. If $\AA$ only contains diagonal 
matrices, the algebra is still \abel, and we can construct 
characters by simply taking a particular diagonal entry at a 
particular point on the manifold
\bas
\chi(a) = a(p)_{ii},\quad a\in \AA,~ p \in M,~ i=1,~\ldots,n.
\eas 
The distance between two points $p$ and $q$ on two copies 
labeled by 
$i$ and $j$ is now given with the help of the characters 
\bas
\chi(a) = \tr(H^i a) (p) \quad \mbox{~and~} \quad 
\psi(a) = \tr(H^j a) (q),
\eas 
where $(H^i)_{nm} = \delta^i_n \delta_{nm} $, as
\ba
d(\chi,\psi) =  \sup_{a \in {\cal A}  } \Big\{  | \chi(a) - 
\psi(a)  | ; ~~ \parallel  [{\cal D}, a  ]  
\parallel \leq 1  \Big\}. 
\label{metriikka1}
\ea

The matrices $H^i$ are Cartan elements of the Lie algebra 
${\bf u}(n)$ in the fundamental representation. 
The distances between the p-branes according to   
\eq~(\ref{metriikka1}) seem to be associated to root 
vectors of 
${\bf su}(n)$. The ${\bf u}(n)$ Cartan element proportional 
to unity ${\unity}_n$ is then naturally  associated to the 
movement of the center of mass described by ${\bf u}(1) 
\subset{\bf u}(n) $.  In the case of 
Lie-algebrae ${\bf g} = {\bf sp}(2r)$ and $ {\bf so}(2r)$ 
the diagonal elements are of the form 
${\rm diag}(A,-A^{T})$. 
Each p-brane appears thus twice: one might say that there is 
a mirror in the noncommutative space.
We now have characters 
\bas 
\chi(a) = \tr (P_{\pm} H^i a) (p),
\eas 
where $P_+ = {\rm diag} ({\unity}_r,0)$ and
$P_- = {\rm diag} (0,{\unity}_r)$ are projections 
to different sides of
the mirror, and $H^i \in {\bf h}$ is a Cartan element of the
Lie-algebra ${\bf g}$.  In this way one may try 
to extend ${\bf g}$ to the exceptional
Lie algebrae and to the Ka\v{c}--Moody algebrae. 
The latter possibility
might enable one to include even the winding modes 
of string theory into the NCG approach.

In the direct \nabel~generalization the functions $\chi$
introduced above fail to be homomorphisms.  Let us, however,
{\em postulate} that those functions 
$\chi: \AA \rightarrow {\Cset}$ that reduce to characters 
in the diagonal subalgebra $\AA_0 \subset \AA$ 
are to be associated with distances in the genuinely
\nabel~NCG. The distances then depend 
on the choice of the Cartan subalgebra, and thus 
explicitly break global gauge invariance. This is actually 
natural, since the global gauge invariance here is the 
counterpart of 
the Lorentz invariance of the
compactified space,  also broken by the introduced p-branes.

\subsection{Differential geometry in NCG}

In order to address questions that concern geometry, 
we shall need the generalized differential forms of NCG: 
The differential algebra of forms over a 
noncommutative space is given
in terms of the tensor algebra of $\AA$. 
A p-form $\alpha = a^0 {\rm
d} a^1 \ldots {\rm d} a^p \in \Omega^*\AA$ is 
represented as an
operator\footnote{Since $\pi$ is assumed a 
faithful representation of
$\AA$ on $\HH$ we write $\pi(a) \equiv a,~ a \in \AA$.}  
on $\HH$ by
\bas \pi(\alpha) = a^0 [\DD, a^1] \ldots [\DD, a^p].  
\eas However, to
avoid the case where $\pi(\alpha) = 0$ but 
$ \pi({\rm d}\alpha) \neq
0$ one should properly consider the equivalence classes \bas
\pi_D(\alpha) \in \Omega_{D}^*\AA = \pi [ \Omega^*\AA / 
(\ker \pi +
{\rm d} \ker \pi) ].  \eas The concept of a fiber bundle 
has its
generalization in NCG, as well.  Here we only need the 
trivial bundle
over $\AA$ that has a connection $\dd + \rho$, 
where $\rho \in
\Omega^1\AA$, and a curvature $\vartheta = {\rm d} 
\rho + \rho^2 \in
\Omega^2\AA $. We choose $\rho^* = -\rho $ so that 
$\vartheta^* = \vartheta$.  
Such a connection 1-form can be given as a (formal) sum
\ba \rho = \sum a^n {\rm d} b^n, \label{roo} \ea where the 
sequence
$(a^n , b^n)$ is chosen to satisfy $\sum a^n b^n = {\unity}$.

We can define, under suitable conditions, an inner product 
$ \langle ~ | ~ \rangle $ in $ \Omega^*\AA $ 
by setting 
\ba \langle \alpha | \beta
\rangle = {\rm Tr}^+(\pi (\alpha^* \beta) | \DD|^{-p}) ,
\label{normi}
\ea 
where the trace is as in \cite{varilly} and $p = \dim M$.  
This
can be used to project $\Omega^*\AA \rightarrow 
\Omega^*_D\AA $ by
choosing a representative of $\alpha$ in $\Omega^*_D\AA$ 
such that it
minimizes the norm $ \langle \alpha | \alpha \rangle$. 

The NCG-version of a Yang--Mills action is 
\bas S_{YM}(\rho) = \langle
\vartheta(\rho) | \vartheta(\rho) \rangle_D = {\rm
Tr}^+(\pi_{D}(\vartheta(\rho))^2 | \DD|^{-p}) .  
\eas 
This can be evaluated further in the cases that 
we shall consider: The result is 
\bas
S_{YM}(\rho) = C_p \int_{M} {\rm tr} 
(\pi_{D}(\vartheta (\rho))^2),
\eas 
where $C_p$ is a dimension dependent constant and 
${\rm tr}$ is taken over the finite matrix indices.

\section{The construction}

Our aim is to construct a NCG that 
produces a Yang--Mills--Higgs
theory with a scalar potential and a fermionic 
sector that match those
of a dimensionally reduced Yang--Mills theory. 
It turns out to be necessary to consider mass matrices 
that are a tensor
product of a Clifford algebra and an arbitrary matrix 
algebra. This also automatically leads to 
the right fermion sector. 
Our construction is essentially a generalization of 
\cite{cham_susy} where four-dimensional supersymmetric 
field theories where studied in NCG.

Let $M_p$ be a  $p+1$ dimensional compact spin-manifold 
with a Euclidean
 metric  $g_{\mu\nu}$ and  $p$ odd. The \herm~generators 
of its Clifford algebra are $\gamma^{\mu}$,  
and the \herm~chirality 
operator $\gamma_{p+2}$ satisfies 
$\gamma_{p+2}^2 = {\unity} $. The 
spin-connection $D_{\mu}$
operates on the square integrable spinors in $ L^2(S(M_p))$.
Let further $\Sigma^a, a=1,\ldots, \tilde{p}$ be 
$s \times s$-matrix-valued functions that satisfy 
\bas 
\{ \Sigma^a, \Sigma^b \} =  2 g^{ab}, 
\eas
where $g^{ab}$  are scalar functions on  $M_p$. 
The matrix $(g^{ab}) >0 $ has an inverse $(g_{ab})$. 
These two algebrae can be naturally  combined into a 
Clifford algebra with generators
\bas
\Gamma^{\mu} &=& \gamma^{\mu} \otimes {\unity}, ~~ \mu = 
0,\ldots,p \\
\Gamma^{a+p} &=&  \gamma_{p+2} \otimes \Sigma^a,~~ a= 1,\ldots, 
\tilde{p}.
\eas
Also this Clifford algebra has a \herm~chirality 
operator for even $\tilde{p}$
\bas
\Gamma_{D+1} = \gamma_{p+2} \otimes \Sigma_{\tilde{p}+1},
\quad D = \tilde{p} + p + 1.
\eas 

The NCG's we are interested in are given by the 
K-cycle $(\HH,\DD,\Gamma)$ on the *-algebra $\AA$ 
\ba
\AA &=& C^{\infty}(M_p, {\Cset}) \otimes U({\bf g}) \\
\HH &=&  L^2(S(M_p))\otimes {\Cset}^s \otimes  L \otimes 
{\Cset}^k \\
\DD &=& \Gamma^{\mu} D_{\mu} \otimes {\unity}_{ L } \otimes 
{\unity}_k + \Gamma^a \otimes S_a \otimes K \label{dirr} \\
\Gamma &=& \Gamma_{D+1}  \otimes {\unity}_L \otimes {\unity}_k.
\ea
Here $U$ denotes the universal enveloping algebra of a Lie 
algebra, $S_a \in \II$, $\II  \subset {\bf g}$ is a (finite) 
subset containing 
\aherm~elements of a Lie-algebra ${\bf g}$ 
of a Lie-group $G$ and the \herm~matrix 
$K$ mixes the $k$ fermion 
flavours included in the Hilbert 
space. Notice that $U({\bf u}(n)) = M_n({\Cset}) $ in the 
fundamental representation. 
The operator  $\DD$ is \aherm~on $\HH$, and 
 it anticommutes with $\Gamma$.
The algebra $\AA$ acts in $\HH$ through the (irreducible) 
representation $R$ in the $n$-dimensional space $L$ as
\bas
\pi: \AA \rightarrow {\rm End}  (\HH) ; ~ a \mapsto  
{\unity}_{S(M_p)} \otimes {\unity}_s \otimes 
{R(a)}  \otimes {\unity}_k.
\eas
In the following  all tensor products, unit matrices 
and explicit summations as in \eq~(\ref{roo}) will 
be omitted.

\subsection{The NCG dimensional reduction}

The geometric content of the theory can be elucidated 
by considering differentials 
\bas
\pi(\dd a) = \FMslash{\partial} a + 
\Gamma^a \Big( (S_a, \alpha) a^{{\alpha}} E^{{\alpha}} - 
S_a^{{\alpha}}  (a,\alpha) E^{{\alpha}} + 
S_a^{{\alpha}} a^{{\beta}} \varepsilon_{\alpha, \beta} 
E^{{\alpha +\beta}}   +
S_a^{{\alpha}} a^{-{\alpha}} H^{\alpha}
\Big)
\eas 
in the Cartan--Weyl basis.
Summation is assumed over repeated indices and $(,)$ 
is the inner product in the root space. 
Consider in particular the case 
\bas
S_a = S_a^{\alpha} (E^{\alpha}- E^{-\alpha}),
\eas 
where the roots belong to the lattice $\Phi$ 
of  ${\bf su}(n)$.
Let  $a \in \AA_0$ be diagonal and denote $\alpha 
= {\bf e}^i - {\bf e}^j$. 
A differential of $\pi(a) $ becomes
\bas
\pi(\dd a) &=& \Gamma^{\mu} \partial_{\mu} a^i H^i + 
\Gamma^a S_a^{\alpha} (a^j - a^i) (E^{\alpha} + E^{-\alpha}). 
\eas
One can view $i,j$ as 
labels of two lattice points a distance $\epsilon_a$ 
apart from each other where 
\bas
S_a^{\alpha} = \frac{ \epsilon_a }{  \epsilon_a ^2 } ~ 
\delta_a^{\alpha} .
\eas 
Taking the  limit $\epsilon \rightarrow 0$ with 
$a^i -  a^j = {\cal O}(\epsilon)$ one 
reduces the differential to the standard form
\bas
\vec{a}(x + h) - \vec{a}(x) \equiv \tr_{{\rm Cliff}} \Big( 
\FMslash{h} \pi(\dd a) \Big)
 = Da \cdot h + {\cal O}(h)^2
\eas
where $ h=(h_{\mu}, \epsilon_a )$ and the arrows refer to a 
basis of the vector space ${\bf g}$. 
Moving from the diagonal element $a^i$ 
to the element $a^j$ has thus an interpretation as motion 
in a larger space to a direction corresponding to the root 
${\bf e}^i - {\bf e}^j$. Depending on the choice of the
matrices $S_a$ there is an index $a = 1,\ldots,\tilde{p}$ 
that corresponds to this direction. The only novelty 
here is that the vector space ${\bf g}$'s basis elements 
 do not commute. The theory is thus indeed 
$D$ dimensional: the additional $\tilde{p}$  
directions appear in a complicated way in the mutual 
relations of the p-branes.  

\subsection{Distances}

Distances calculated from \eq~(\ref{metriikka1})  
depend on the 
choice of the Cartan subalgebra. The simplest case that will 
turn out to be interesting from the field theory point of view 
as well, is $\II \subset {\bf h}$ in which case all distances 
become infinite. 

Let us next conjugate $\II$ by
\bas
g = \exp \Big( \frac{i}{2}  \sum_j \varphi_j 
(E^{ \alpha^{(j)} } + E^{ - \alpha^{(j)} }) \Big) 
\in {\rm SU}(n),
\eas
where the roots for simplicity satisfy the condition
\bas
({\bf \alpha}^{(i)*},  {\bf \alpha}^{(j)}) = 2 \delta^{ij}.
\eas
The new Dirac operator is 
\bas \DD = \Gamma^{\mu} D_{\mu}  + 
\Gamma^a ( {\rm Ad}_g  S_a)   K.
\eas 
A nontrivial lower bound for the distance between a pair of 
p-branes corresponding to a root $\alpha \in \Phi$ 
is obtained by first noticing that $\tr(H^{\alpha} a)$ 
in \eq~(\ref{metriikka1}) only depends on the part 
of $a$ proportional to $H^{\alpha}$. One then estimates 
the constraint term from above, 
and saturates the estimate by choosing 
$H^{\alpha} a_0 = a$ for some number $a_0 \in \Cset$.
We then get
\bas
d(\alpha) \geq \left( \max_j ~ \Big\{ m_j ~
| (\alpha^*, \sin \varphi_j ~\alpha^{(j)}) | 
\Big\} \right)^{-1},
\eas 
up to a constant normalization, where 
\ba 
m^2_j = 2 g^{ab} (\alpha^{(j)}, {S}_a)  
(\alpha^{(j)},  {S}_b) \label{massa}
\ea
is the  $W$-boson mass in a theory  
in which the adjoint representation scalars get vev's 
$\langle \Phi_a \rangle = S_a$ and $g_{ab}$ 
is a metric in the space of the scalars $\Phi^a$. 
This will happen
in the theory at hand as well, and thus unbroken 
gauge symmetry $(S_a,\alpha) = 0$ implies infinite 
distance.  

Notice that here (the estimate of) the metric 
$d$ does not necessarily 
satisfy the triangle inequality. 
The situation with D-branes is exactly the opposite: 
$d_D(\alpha) = \alpha' m_{\alpha}$, unbroken gauge symmetry 
implies zero distance and the triangle inequality is valid. 

\section{The field theory}

We are now ready to find out to which Yang--Mills 
theory the present NCG model gives rise.
Let us choose an \aherm~1-form $\rho \in 
\Omega^1 \AA$ as in (\ref{roo}) and write  
\bas
\pi(\rho) = \Gamma^{\mu} A_{\mu} + \Gamma^a(A_a - S_a), 
\eas
where
\bas
A_{\mu} &=& a D_{\mu} b = a \partial_{\mu}b \\
A_c &=& a [ S_c , b ]  + S_c =  a S_c b.
\eas
The curvature becomes
\bas
\pi(\vartheta) &=& X + (g^{ab} A_a A_b - Z) K^2 + 
\frac{1}{2}(F_{\mu\nu} - T_{\mu\nu}^{\kappa} A_{\kappa}) 
\Gamma^{\mu\nu} \nonumber  \\
 & & + ( D_{\mu} A_a ) \Gamma^{\mu a} K + 
\frac{1}{2} 
([A_a,A_b] - f_{ab}^c A_{c} - \tilde{f}_{ab}^c Y_{c})  
\Gamma ^{ab} K^2,
\eas
where $F_{\mu\nu}$ is the field strength of $A_{\mu}$, 
$T_{\mu\nu}^{\kappa}$ are the structure constants 
of the algebra spun by $\partial_{\mu}$, $f_{ab}^c $ and 
$\tilde{f}_{ab}^c $  are the ${\bf g}$-structure 
constants and the spin-connection has become a 
${\rm U}(n)$-covariant connection $D_{\mu}  \rightarrow 
D_{\mu} +[A_{\mu},~~] $. Functions  
$X$  and $Y_c$ are independent of $A_B$, where $B = (\mu,c)$. 
$Y_c$'s Lie-algebra index $c$ refers to 
the subspace of ${\bf g}$  orthogonal to ${\rm sp } ~\II$ 
under the Killing metric. 
The function $Z = a C b$, where $C = g^{ab} S_a S_b$, 
is generically a free field.

If $K = {\unity}$ the ${\bf u}(1)$-part of $A_{B}$ couples 
only to fermions, as we shall see, and decouples completely 
if the fermions are 
in the adjoint representation of ${\bf g}$. As was shown 
before, 
this part of the gauge group should be associated to the 
motion of the center of mass. We impose for simplicity 
the constraint $\tr ~ A_B = 0$.
We shall also assume $T=0$.

The next problem is to find a representative of 
$\vartheta$ in $\Omega^2_D\AA$. The 1-forms $\sigma \in 
\ker \pi $ give rise to those 2-forms $\dd \sigma$ 
that constitute the 
ambiguity in the choice of this representative. 
The ambiguity is actually just the freedom to shift the 
\nondyn~fields at will, and choosing the representative of 
$\vartheta$ as suggested above amounts to
eliminating the \nondyn~fields $X, Y_c$ and, depending 
on the choice of $\II$, also $Z$ by imposing their classical
equations of motion. 

Suppose the matrix $C$ is expressible as 
a linear combination 
\bas
C = \tr C ~{\unity}_n + 2 C^a~ S_a.
\eas 
Then the field $Z$ is 
not free, and we only need to eliminate $X$ and $Y_c$. The 
resulting action is 
\bas
S_{YM}  &=&  
 - \half  \int_{M_p} \Bigg( \tr F_{\mu\nu}^2 
 + 2 \tr (D_{\mu} A_a)^2  -  2 \kappa~ 
 \tr \Big( (A_a - C_a)^2 - C_a^2 - \tr C \Big)^2 \nonumber \\
 & &  \qquad \qquad  + (1 + \kappa)~ \tr \Big( (
 [A_a,A_b] - 
f_{ab}^c A_{c}) P_{ab,de}^{\perp} \Big)^2 ~ \Bigg) 
\label{vaik}, 
\eas
where $\kappa = \tr ~ K^4$, $K$ is normalized to 
$\tr ~ K^2 = 1$ and $P^{\perp}$ is a projection 
to the subspace of $v\wedge w \in {\rm sp} 
\II \wedge {\rm sp} \II $ with the property $ [v,w] 
\in {\rm sp} \II$.
If $C$ is not of the form suggested above then 
the third term in \eq~(\ref{vaik}) vanishes.

Choosing $K = {\unity}$ 
and 
$ [\II,\II] = {\bf 0} $
the Yang--Mills action reduces to
\ba
S_{YM} 
 =  -\half  \int_{M_p} \tr (  F_{\mu\nu}^2 
 + 2 (D_{\mu} A_a)^2  +  [A_a,A_b]^2)   
 =  - \half  \int_{M_p} \tr  F_{AB}^2. \label{red}
\ea
This is the trivially from $D$ dimensions 
down to $p+1$ dimensions reduced Yang--Mills theory.  

The dynamics of the fermions $| \psi \rangle 
\in \HH$  is determined by
the action
\ba
S_F &=& ~ \langle \psi |    
\DD + \pi_D(\rho)  | \psi \rangle  \label{fermi} 
= \int_{M_p} \tr ~ \bar{\psi} \Gamma^B D_B \psi. 
\ea
 Under assumptions that 
the elements of $\II$ commute and that the gauge fields be 
traceless  this theory describes the $D$ dimensional 
${\rm SU}(n)$ Yang--Mills coupled to  fermions after the 
trivial dimensional reduction to $p+1$ 
dimensions. This is an 
immediate consequence of the structure of the Dirac operator
$\DD$.

In particular, if $D=10$ and $\HH$ contains Majorana--Weyl 
fermions or $D=6$ and $\HH$ 
contains Weyl fermions, we get in 
$p+1=4$ dimensions N=4 and N=2 
supersymmetric Yang--Mills 
theories, respectively \cite{cham_susy}. 
For this, we let the covariant 
derivative  $\pi_D(\dd + \rho)$ act on the ${\bf g}$-valued 
fermion fields in (\ref{fermi}) through the Lie-brackets.

\section{Symmetries and classical moduli}

The theory is invariant under unitary transformations 
$u \in {\cal U} (\AA)$ 
\bas
\rho & \rightarrow & u \rho u^*   + u \dd u^*, \quad \rho 
\in \Omega^1\AA \\
| \psi \rangle  & \rightarrow & R(u) | \psi \rangle , \quad   | 
\psi \rangle \in \HH.
\eas
In terms of its constituent fields $a^m, b^m \in \AA$ of 
\eq~(\ref{roo}) the 
transformation of the connection 1-form 
is expressible as  $(a^m,b^m) 
\rightarrow (u a^m, b^m u^*) $. 
This symmetry gives rise to the local ${\rm SU} (n)$ gauge 
symmetry.

By a global gauge transformation one usually means 
a $u \in {\cal U} (\AA)$ that satisfies 
$\dd u = 0$. In the present case $u$ would thus be a 
constant matrix that commutes with $\II$. 
However, let us relax this 
condition and consider such transformations 
of the algebra $\AA$ that become global 
symmetry transformations of the field theory, 
\ie~$u \in {\rm Int}~ G $. 
On the level of the differential algebra and 
the choice of the K-cycle these transformations act as
\bas
\omega &\rightarrow & {\rm Ad}_u \omega , \quad \omega \in 
\Omega^*\AA \\
\DD &\rightarrow &  {\rm Ad}_{u^*}  \DD= 
\Gamma^{\mu} D_{\mu}   + \Gamma^a \otimes  {\rm Ad}_{u^*} S_a.
\eas
A global gauge transformation is thus essentially 
a change of NCG. There is consequently
a whole orbit of NCG's that yield the same field 
theory. Notice, however, that since the formula for 
distances 
is not $G$-invariant, the distances between the branes vary 
as we move along the orbit of $G$ in the parameter space.

Let us consider the Yang--Mills 
theory of \eq~(\ref{red}). The 
vacuum expectation values $\langle A_a \rangle $ that 
minimize the potential $ - \tr [A_a, A_b]^2 $ 
commute. Thus for any dimensionally reduced Yang--Mills 
 theory with vev's $\langle A_a \rangle = S_a$ 
there is a NCG of the form suggested above with the property 
$A_a = a [S_a,b] + S_a = S_a$ 
for the  configuration $a=b={\unity}$. This can be inverted by
simply {\em  postulating} that the coefficients $S_a$ { 
are} the vacuum expectation values of the fields $A_a$ and 
that the vacuum corresponds to the configuration 
$a=b={\unity}$. This is the point of view also adopted in the 
study of symmetry break down in Ref.~\cite{froeh_gut}. The 
moduli space of vacua thus becomes the moduli space of 
NCG's. 

For ${\bf g} = {\bf su}(n)$, the 
moduli space of the considered NCG's consists of vectors 
\ba
(S_a) \in {\cal M}_{NCG} = {\rm Ad}_G ( {\bf h^{n-1}})/{ W}, 
a=1,\ldots,n-1, \label{modd}
\ea 
where the group $W$  acts by permutations in the index $a$. 
In the fixed points of $W$ we get degeneracy in $S_a$. This 
means that we can choose a new basis of $\Gamma^a$'s so 
that some of the new coefficients $S_a$ vanish in 
\eq~(\ref{dirr}). This 
leads to $A_a \equiv 0$ for some $a$, and to the restoration 
of some of the gauge symmetry. 

In the \abel~limit $\AA \rightarrow
 \AA_0$ the gauge symmetry becomes a 
local ${\rm U}(1)^{n-1}$ and the corresponding gauge fields 
$A_{\mu}^i$ decouple. Due to the condition $ab = {\unity}$ 
we get $A_a = S_a$, if the $S_a$'s are diagonal, and the 
$W^{\pm \alpha}$-bosons would 
get a mass $m_{\alpha}$ of 
\eq~(\ref{massa}) if there were any. Even in the case 
that the $S_a$'s are not diagonal, the vector bosons 
keep out of the 
theory, and one sees fluctuation fields around the vev's 
$\langle A_a \rangle = S_a$ only in those 
directions of the matrix space, 
where the $S_a$'s have components. 

\section{Conclusions} 

We have analysed a subclass of NCG's that describe a collection 
p-branes with $p$ odd. The studied NCG's give rise
to field theories that can also be obtained by dimensional 
reduction. The additional dimensions enter the 
NCG formulation in the form of the p-branes' mutual 
relations. 

The metric of the noncommutative space turned out not to be 
uniquely determined by the commuting analogue. The latitude in 
its definition was related to the mutual orientations of the 
geometric objects. In addition, the usual 
matrix structure of the NCG models was 
obtained {}from an underlying group structure  
intimately connected with the measurement of distances. 
The classical moduli of the field theory could also be related 
to the parameters of the NCG models in a transparent manner. 

The obtained field theories also describe 
low energy physics of D-branes. Despite the similarities 
of the presented models to the D-brane effective 
theories, there are differences \cite{douglas}:
Most importantly, the distances one obtains 
between p-branes are here of the form $1/m_{\alpha}$ 
whereas in string theory one obtains $\alpha' m_{\alpha}$. 
Also, it is not known, how to give correct dynamics 
to the metric $g_{AB}$, how to incorporate winding 
modes of strings or how to extend the theory to 
higher orders in the field strength.

{\bf Acknowledgements}: 
We thank  A.J.~Niemi for discussions.

\end{document}